\documentclass[amstex,12pt]{article}
\usepackage{amsmath}
\usepackage{amsfonts}
\usepackage{amssymb}

\newtheorem{theorem}{Theorem}

\newtheorem{definition}[theorem]{Definition}

\input epsf
\epsfverbosetrue
\newtheorem{figrpic}{Figure}

\def\FRAME1#1#2#3#4#5#6#7#8
{
 \begin{figrpic}
 \epsfysize=#3{\centerline{\epsfbox{#7}}}
    {\begin{center}
      {\rm {\ #5} }
     \end{center}
    }
 \label{#6}
 \end{figrpic}
}
\input{tcilatex}

\begin{document}

\title{Lobby index in networks}
\author{A. Korn\thanks{%
Department of Telecommunications and Media Informatics, University of
Technology and Economics Budapest,\ 1117 Budapest, Magyar tud\'{o}sok k\"{o}r%
\'{u}tja 2, korn.andras@tmit.bme.hu}, A. Schubert\thanks{%
Institute for Research Organisation, H-1245 Budapest, P.O.Box 994,
schuba@iif.hu}, A. Telcs \thanks{{Department of Computer Science and
Information Theory, \ University of Technology and Economics Budapest , \
1117 Budapest, Magyar tud\'{o}sok k\"{o}r\'{u}tja 2, telcs@szit.bme.hu} }}
\maketitle

\begin{abstract}
We propose a new node centrality measure in networks, the lobby index, which
is inspired by Hirsch's $h$-index. It is shown that in scale free networks
with exponent $\alpha $ the distribution of the $l$-index has power tail
with exponent $\alpha \left( \alpha +1\right) $. Properties of the $l$-index
and extensions are discussed. \newline
\end{abstract}

Efficient communication means high impact (wide access or high reach) and
low cost. This goal is common in communication networks, in society and in
biological systems. In the course of time many centrality measures proposed
to characterize a node's role, position, or influence in a network but none
of them capture the efficiency of communication. \ This paper is intended to
fill this gap and propose a new centrality measure, the lobby index.

Hirsch \cite{H} proposed the $h$-index: \emph{"the number of papers with
citation number }$\geq h$\emph{, as a useful index to characterize the
scientific output of a researcher". }Barab\'{a}si \& al \cite{BAJ} devised a
very simple network model which has several key properties: most importantly 
\emph{the degree distribution has a power-law upper tail, the node degrees
are independent, and typical nodes are close to each other. }Schubert \cite%
{S0} used the $h$-index as a network indicator, particularly in scale-free
networks. This paper is devoted to the characterization of network nodes
with a $h$-index type measure.

\begin{definition}
The $l$-index or \textbf{lobby index} of a node $x$ is the largest integer $%
k $ such that $x$ has at least $k$ neighbors with a degree of at least $k$.
(See also $\left( \ref{ldef}\right) \,$.)
\end{definition}

In what follows some properties of the lobby index are investigated; it is
shown that in \textbf{S}cale \textbf{F}ree (SF) networks, with exponent $%
\alpha $, the distribution of the $l$-index has a fat tail with exponent $%
\alpha \left( \alpha +1\right) $. Furthermore the empirical distribution of
the $l$-index in generated and real life networks is investigated and some
further extensions are discussed.

\section{Centrality measures}

\emph{\ }Freeman's prominent paper \cite{F} (1979) pointed out that: "\emph{%
Over the years, a great many measures of centrality have been proposed. The
several measures are\ often only vaguely related to the intuitive ideas they
purport to index, and many are so complex that it is difficult or impossible
to discover what, if anything, they are measuring.}" It is perhaps worth
noting that research in this field dates back to Bavelas \cite{Ba} (1949). \ 

At the time of Freeman's paper most centrality measures were equivalents or
modifications of the three major and widely accepted indexes, the degree (cf.%
\cite{Sh}), closeness (cf. \cite{Ba1} and \cite{Sa}) and betweenness (cf. 
\cite{Ba}) centrality\ (see also Borgatti \cite{Bo}).

As time passed, many new centrality measures were proposed. After years of
research and application, the above three and eigenvector centrality (a
variant of which computer scientists call PageRank \cite{BP}\ and Google
uses to rank search results) can be said to have become a standard; the
others are not widely used. The historical three and eigenvector centrality
are thus the conceptual base for investigating centrality behavior of nodes
and full networks.

Notwithstanding Freeman's wise warning the present paper proposes the lobby
centrality (index) in the belief that Hirsch's insight into publication
activity (which produces the citation network) has an interesting and
relevant message to network analysis in general.

\emph{The diplomat's dilemma} It is clear that a person has strong lobby
power, the ability to influence people's opinions, if he or she has many
highly connected neighbors. This is exactly the aim of a lobbyist or a
diplomat \cite{HG}. The diplomat's goal is to have strong influence on the
community while keeping the number of his connections (which have a cost)
low. \ If $x$ has a high lobby index, then the $l$-core $L\left( x\right) $
(those neighbors which provide the index) has high connectivity
(statistically higher than $l\left( x\right) $, see $\left( \ref{trunc}%
\right) $ and the comment there). In this sense, the l-index is closely
related to the solution of the diplomat's dilemma.

\emph{Communication networks }Research of communication networks and network
topology are in interaction. Node centrality measures are essential in the
study of net mining \cite{n1}, malware detection \cite{n2}, in reputation
based peer to peer systems \cite{n3}, delay tolerant networks \cite{n4} and
others (see \cite{n5}, \cite{n6} and the references therein). We expect that
in the case of social and communication networks (some of which are also
based on social networks) the lobby-index is located between the bridgeness 
\cite{NPNB}, closeness, eigenvector and betweenness centrality. Based on
this intermediate position of the lobby index we expect that it can be a
useful aid in developing good defence and immunization strategies for peer
to peer networks as well as help create more efficient broadcasting schemes
in sensor networks and marketing or opinion shaping strategies.

\emph{The distribution of the }$l$\emph{-index }Let us consider scale-free
networks and assume that the node degrees are independent. \ The degree is
denoted by $\deg \left( x\right) $ for nodes and the $l$-index is defined as
follows. Let us consider all $y_{i}$ neighbors of $x$ so that $\deg \left(
y_{1}\right) \geq \deg \left( y_{2}\right) ...$; then, 
\begin{equation}
l\left( x\right) =\max \left\{ k:\deg \left( y_{k}\right) \geq k\right\} .
\label{ldef}
\end{equation}

\textbf{Theorem}

If the vertex degrees are independent and ${\mathbb{P}}\left( \deg \left(
x\right) \geq k\right) \approx ck^{-\alpha }$ for all nodes $x$, then

\begin{equation}
\fbox{${\mathbb{P}}\left( l\left( x\right) \geq k\right) \simeq k^{-\alpha
\left( \alpha +1\right) }$}  \label{main}
\end{equation}

for all nodes $x$, \footnote{%
Here and in what follows $a_{n}\approx b_{n}\ $means that $\frac{a_{n}}{b_{n}%
}\rightarrow c$ as $n\rightarrow \infty $ and $a_{n}\simeq b_{n}$ means that
there is a $C>1$ such that for all $n,$ $\frac{1}{C}\leq \frac{\alpha _{n}}{%
b_{n}}\leq C$.}

The proof is provided in the Appendix.

\emph{The Hirsch index}\textbf{\ }The original Hirsch index is based on a
richer model: author $\leftrightarrow $ paper and paper $\leftrightarrow $
citing paper links. Let $x$ be a randomly chosen author of the scientific
community under scrutiny and $n=n\left( x\right) $ is the number of his/her
papers (either in general or within a defined period). Let $y_{i}$ denote
the individual papers (where $i=1,...n$,) and $c\left( y_{i}\right) $ their
citation score (in decreasing order), so that $c\left( y_{1}\right) \geq
c\left( y_{2}\right) \geq ...\geq c\left( y_{n}\right) $. $h\left( x\right) $
is the Hirsch index of $x:$%
\begin{equation*}
h\left( x\right) =\max \left\{ k:c\left( y_{k}\right) \geq k\right\} .
\end{equation*}%
Assume that the paper productivity has an $\alpha $-fat tail: $G_{l}^{P}={%
\mathbb{P}}\left( n\left( x\right) \geq l\right) \approx cl^{-\alpha }$ and
the citation score has a $\beta $-fat tail: 
\begin{equation*}
G_{l}^{Q}={\mathbb{P}}\left( c\left( y\right) \geq l\right) \approx
cl^{-\beta }.
\end{equation*}

Along the lines of the argument that led to $\left( \ref{main}\right) $ one
can see that $h$ has an $\alpha \left( \beta +1\right) $-fat tail \cite{T1}: 
\begin{equation}
\fbox{${\mathbb{P}}\left( h\left( x\right) \geq k\right) \simeq k^{-\alpha
\left( \beta +1\right) }$}  \label{law2}
\end{equation}

\emph{How good is an $l$-index of $k$?} If a node $x$ of degree $n$ has an $%
l $-index of $k$, Gl\"{a}nzel's \cite{G} observation provides a preliminary
assessment of this value:%
\begin{equation}
l\left( x\right) \approx c\deg \left( x\right) ^{\frac{1}{\alpha +1}}
\label{gl}
\end{equation}%
where $\alpha $ is the tail exponent of the degree distribution.
Consequently a lobbyist is doing a good job of solving the diplomat's
dilemma if $l\left( x\right) \gg \deg \left( x\right) ^{\frac{1}{\alpha +1}}$%
. On the other hand our result shows that $l\left( x\right) \geq k$ means
that $x$ belongs to the top $100c_{\alpha }k^{-\alpha \left( \alpha
+1\right) }$ percent of lobbyists.

\emph{The lobby gain} The performance of a lobbyist is indicated by a
measure called the lobby gain. The lobby gain shows how the access to the
network is multiplied using a link to the $l$-core. Let us use the notation $%
D_{i}\left( y\right) =\left\{ z:d\left( x,y\right) =i\right\} $ and set $%
D_{2}^{L}\left( x\right) =\cup _{y\in L\left( x\right) }D_{1}\left( y\right)
\backslash \left[ D_{1}\left( x\right) \cup \left\{ x\right\} \right] $ then
the number of second neighbors reachable via the $l$-core is $\deg
_{2}^{L}\left( x\right) =\left\vert D_{2}^{L}\left( x\right) \right\vert $
and the lobby gain is defined as%
\begin{equation}
\Gamma _{l}\left( x\right) =\frac{\deg _{2}^{L}\left( x\right) }{l\left(
x\right) }.  \label{lg}
\end{equation}%
The lobby gain $\Gamma _{l}\left( x\right) \ $is much larger than one if a
typical link to the $l$-core provides a lot of connections to the rest of
the network for $x$ via that link. It can be shown (see \cite{G}) that the
number of second neighbors reachable via the $l$-core (with multiplicity) is 
$l\left( x\right) ^{2}$.

\emph{The degree distribution within the }$l$\emph{-core\label{core}}\textbf{%
\ }The influential acquaintances of a given lobbyist follow a fat tail
distribution provided the underlying network is SF. In other words if $y\in
L\left( x\right) $ and $l=k$ then\ the truncated distribution (by $k$) of
the degree distribution of $y$ again follows a fat tail distribution: for $%
m>k>0$ 
\begin{equation}
\ {\mathbb{P}}\left( \deg \left( y\right) \geq m|y\in L\left( x\right) \text{
and }l=k\right) \approx c\left( \frac{m}{k}\right) ^{-\alpha }.
\label{trunc}
\end{equation}%
Let us note that this conditional or truncated distribution has a higher
expected value than the original one.

\section{Network examples}

The analysis of different networks received particular attention in the last
decades. The research goals and tools vary greatly. Here we regress to the
roots and consider some "classic" networks and study the distribution of
their lobby index.

\emph{Generated scale-free networks} We have generated 50 20000-node Barab%
\'{a}si (BA) networks \cite{BAJ} with 10 new links each step, starting with
10 initial nodes. \ The degree distribution passed the preliminary test and
has $\alpha =1.96,$ i.e. a $1.96$-fat tail. \ As Figure 1. shows, the
empirical distribution of the lobby index has $\widehat{\eta }=5.14$ while
the theory predicts $\eta =5.76$.

\begin{eqnarray*}
&&\text{The log-log plots for the distribution of l-index } \\
&&%
\begin{array}{cc}
\FRAME{itbpF}{1.8075in}{1.5022in}{0in}{}{}{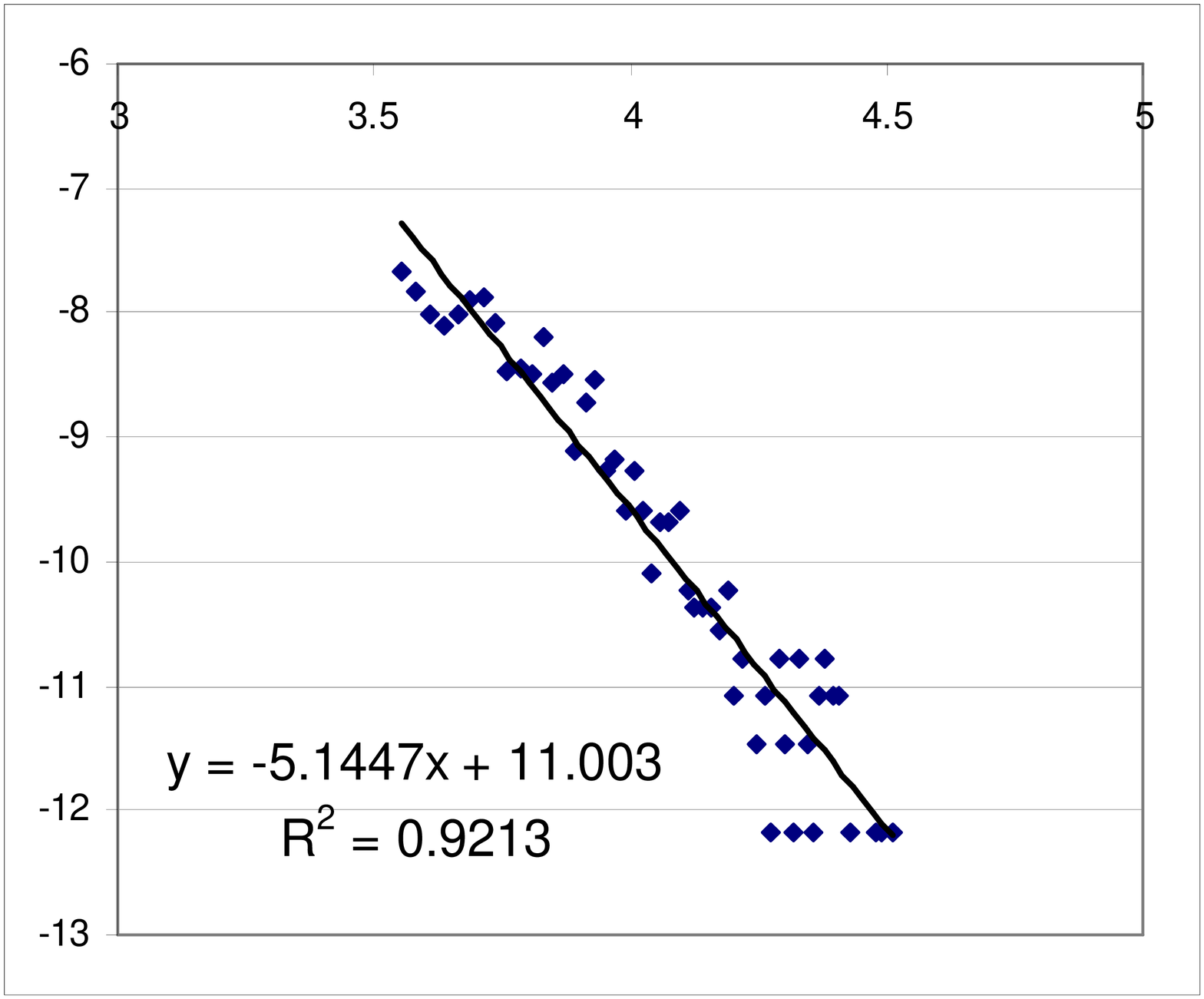}{\special{language
"Scientific Word";type "GRAPHIC";maintain-aspect-ratio TRUE;display
"USEDEF";valid_file "F";width 1.8075in;height 1.5022in;depth
0in;original-width 10.4184in;original-height 8.649in;cropleft "0";croptop
"1";cropright "1";cropbottom "0";filename 'ba1.eps';file-properties
"XNPEU";}} & \FRAME{itbpF}{2.0435in}{1.6224in}{-0.0069in}{}{}{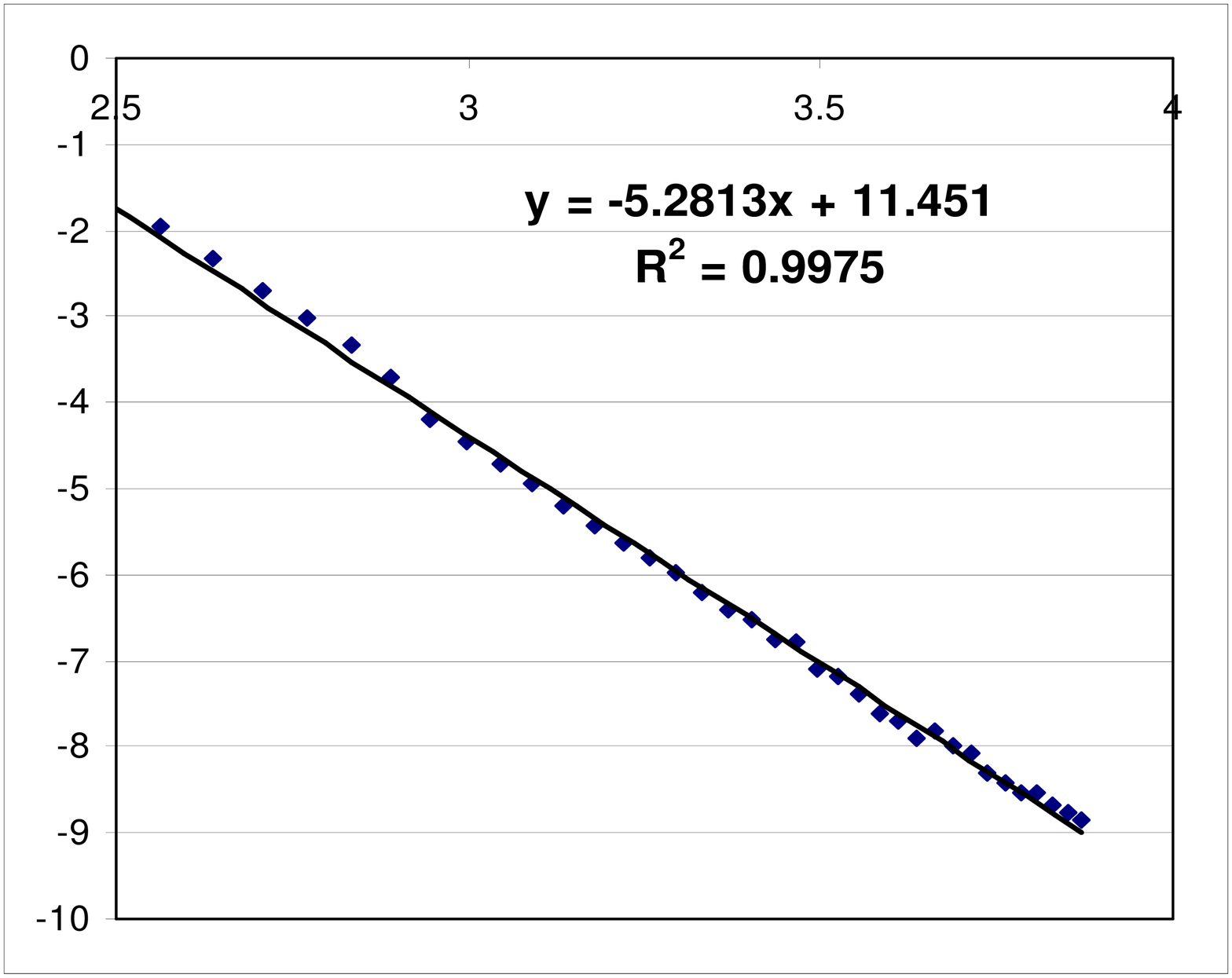}{%
\special{language "Scientific Word";type "GRAPHIC";maintain-aspect-ratio
TRUE;display "USEDEF";valid_file "F";width 2.0435in;height 1.6224in;depth
-0.0069in;original-width 10.4184in;original-height 8.2581in;cropleft
"0";croptop "1";cropright "1";cropbottom "0";filename
'gam1.eps';file-properties "XNPEU";}} \\ 
\FRAME{itbpF}{1.9303in}{1.452in}{0in}{}{}{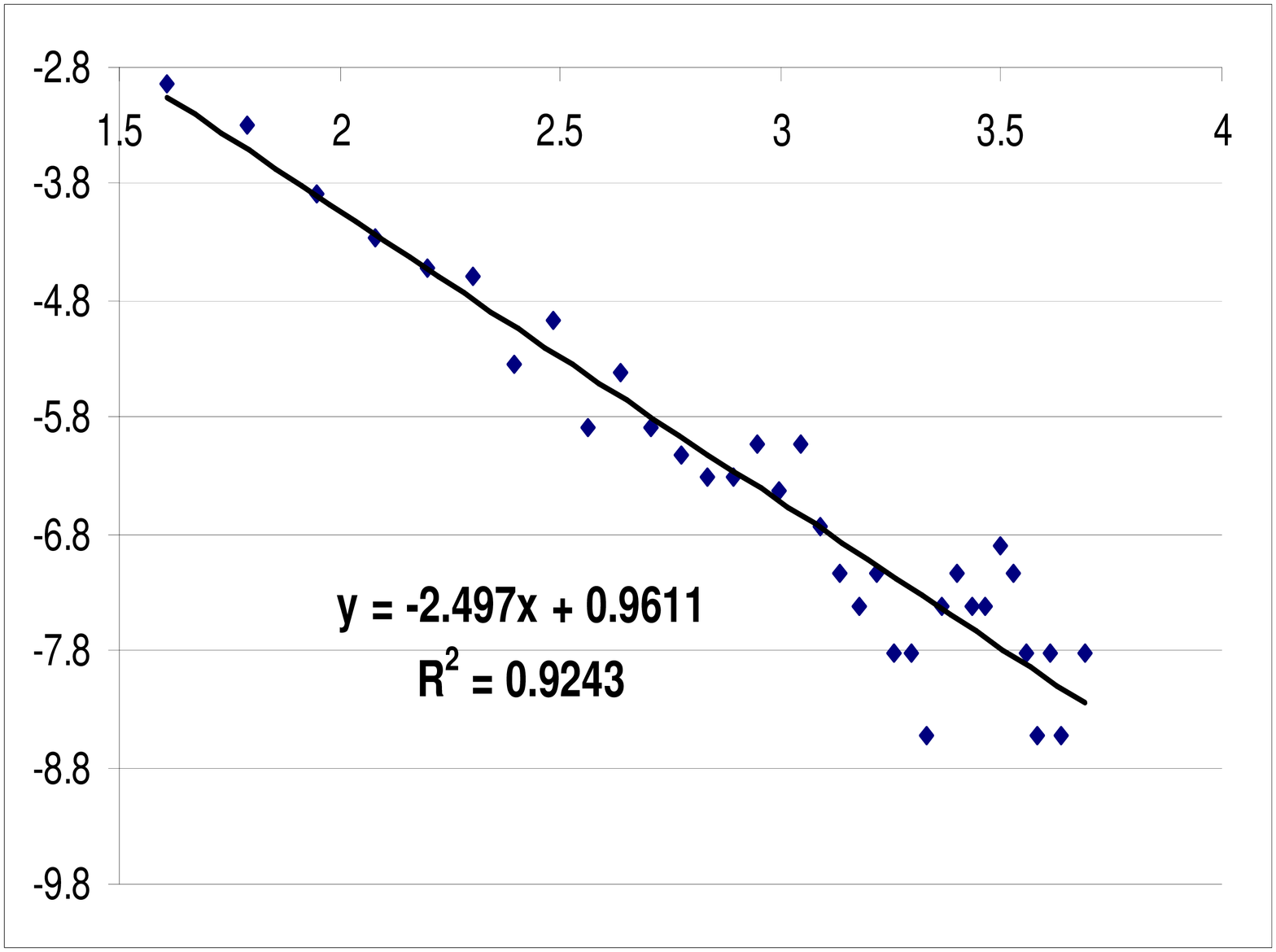}{\special{language
"Scientific Word";type "GRAPHIC";maintain-aspect-ratio TRUE;display
"USEDEF";valid_file "F";width 1.9303in;height 1.452in;depth
0in;original-width 10.5031in;original-height 7.8819in;cropleft "0";croptop
"1";cropright "1";cropbottom "0";filename 'caid1.eps';file-properties
"XNPEU";}} & \FRAME{itbpF}{1.868in}{1.4416in}{0in}{}{}{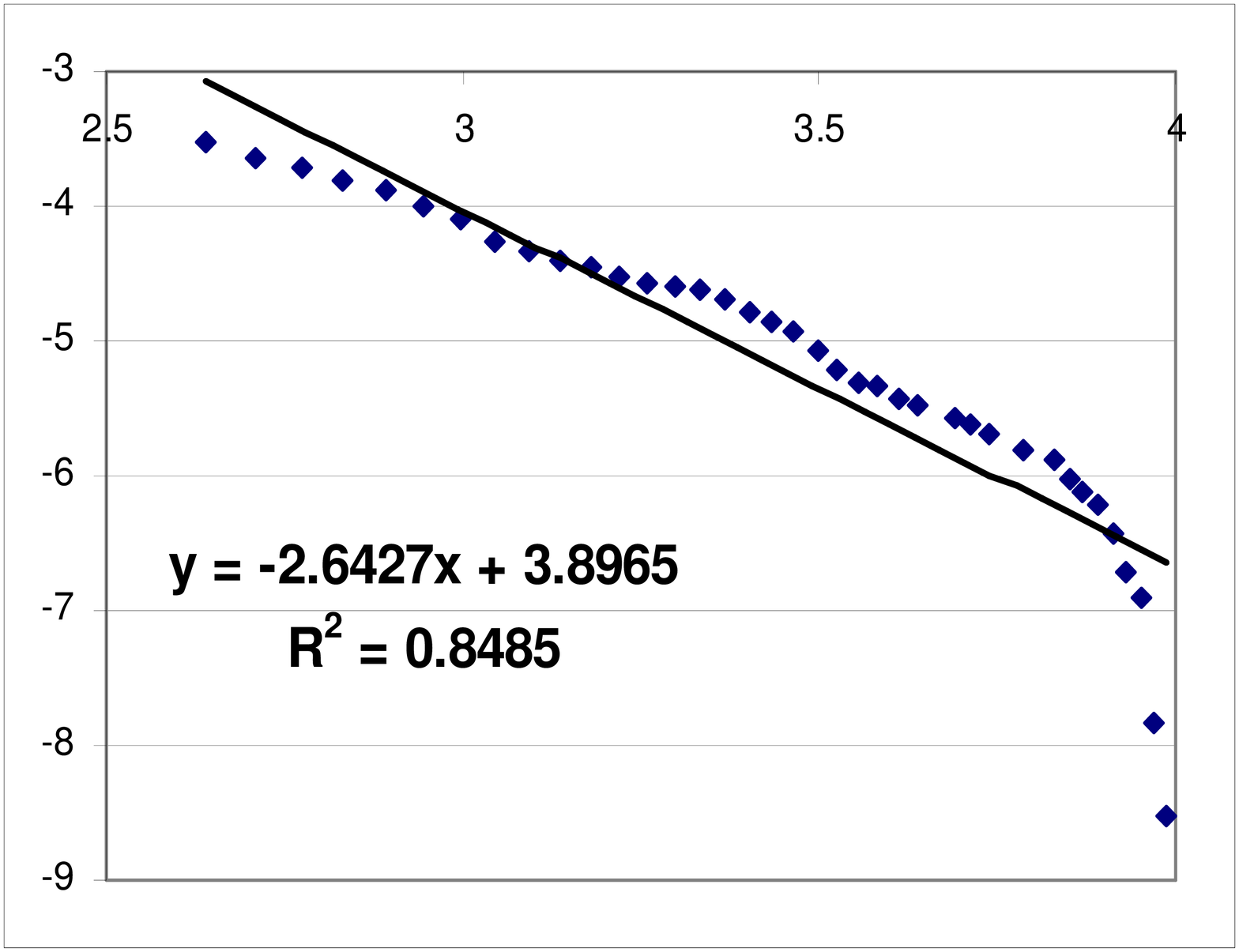}{%
\special{language "Scientific Word";type "GRAPHIC";maintain-aspect-ratio
TRUE;display "USEDEF";valid_file "F";width 1.868in;height 1.4416in;depth
0in;original-width 10.5031in;original-height 8.0929in;cropleft "0";croptop
"1";cropright "1";cropbottom "0";filename 'ig1.eps';file-properties
"XNPEU";}}%
\end{array}%
\end{eqnarray*}%
\bigskip

We have used the generalized Barab\'{a}si model \cite{KRL} which can provide
arbitrary $\alpha >1.$ We generated 50 graphs of $10000$ nodes with the
proposed algorithm and obtained networks with $\alpha =1.9186$ in average,
which would imply $\eta =5.60$; we observed $\widehat{\eta }=5.28.$

Let us remark that the estimate of the tail exponent of fat tail
distributions has a sophisticated technique \cite{CSN} superior to the line
fit on the log-log scale. \ The careful analysis and application of these
methods to the $l$-index will be published elsewhere.

\emph{The AS level graph}\textbf{\ }The \textbf{A}utonomous \textbf{S}ystem (%
\textbf{AS}) level of the Internet infrastructure has already been
investigated in depth (c.f. \cite{ZM} and its bibliography). It turned out
that it not only has a scale-free degree distribution but displays the rich
club phenomenon as well. \ High degree nodes are more densely interlinked
than expected in a \textbf{BA} graph. \ The standard choice for \textbf{AS}
a source of sample data is the CAIDA \cite{CA} project. We determined the
exponent of the tail of the degree distribution and compared it with the
exponent of the tail of the empirical distribution of the l-index. \ We
found that $\alpha =1.61$ and $\eta =\alpha \left( \alpha +1\right) =4.21$
and $\widehat{\eta }=4.14$ is estimated from the empirical distribution of $%
l $.

\emph{The IG model }Mondrag\'{o}n \&al \cite{ZM} proposed a modification of
the Barab\'{a}si network model, the \textbf{I}nteractive \textbf{G}rowth
(IG) model to generate scale-free networks which exhibit the rich club
behavior. In each iteration, a new node is linked to one or two existing
nodes (hosts). \ In the\ first case the host node is connected to two
additional peer nodes using the preferential attachment scheme while in the
latter case only one of the involved (randomly chosen) hosts is connected to
a new peer. We implemented this algorithm and again compared the exponents
extracted from sample data. In this case the network size was 3000 and the
probability of one host was $0.4$ and of two hosts was $0.6$. Again the
log-log fit of the degree distribution tail yielded $\alpha =1.23,\eta =2.74$
and $\widehat{\eta }=2.45$ given by the empirical distribution of $l$.

\emph{The place of the lobby index among other centrality measures} As we
already indicated above the lobby index lies somewhere between the closeness
centrality (\emph{cl}), betweenness (\emph{bw}) and eigenvector (\emph{ev})
centrality. Strong correlation with degree centrality is out of the question
in the light of $\left( \ref{gl}\right) $. In order to gain a better picture
on the behavior of the lobby index we determined the Spearman correlation
between these centrality measures in the AS graph.

\begin{gather*}
\begin{tabular}{||l|l|l|l|l||}
\hline\hline
& l & cl & bw & ev \\ 
l & 1 & $.652$ & $.768$ & $.604$ \\ 
cl & $\mathit{.652}$ & 1 & $.500$ & $.972$ \\ 
bw & $\mathit{.768}$ & $\mathbf{.500}$ & 1 & $.479$ \\ 
ev & $\mathit{.604}$ & $\mathbf{.972}$ & $\mathbf{.479}$ & 1 \\ \hline\hline
\end{tabular}
\\
\text{Table 1. Spearman rank correlation}
\end{gather*}%
The correlations in Table 1. indicate that the $l$-index contains a well
balanced mix of other centrality measures; the $l$-index is slightly closer
to the three "classical" centralities than they are to each other (the
quadratic mean of the three correlations, in boldface, is $0.638$ while the
quadratic mean of of the correlations, in italic, with the $l$-index is $%
0.678$). The Kendall correlations of the investigated centralities have been
calculated and yielded a very similar picture. For biological networks the
Spearman correlation between the closeness and eigenvector centrality is
high (c.f. \cite{KS}); high Pearson correlation can be observed on other
networks as well (c.f. \cite{L}). One centrality measure can be used to
approximate the other, which is not the case with the $l$-index for the AS
graph, but may happen in other types of networks. This will save computation
time given the simplicity of the calculation of the $l$-index. Freeman's
paper \cite{F} as well as \cite{JKS} and \cite{L} are calls for a further
analysis of centrality measures and the $l$-index on different types of
networks.

\emph{Conclusion }A new centrality measure, the $l$-index is proposed and
examined. It is shown that the distribution of the $l$-index has $\alpha
\left( \alpha +1\right) $-fat tail of \textbf{SF} networks with exponent $%
\alpha $. There is a good match between empirical observations (collected in
Table 2) and the theoretical result. \ In this case the aim of the empirical
results was not to verify the theory but to emphasize that the investigated
networks behave in the expected way with respect to the lobby index.%
\begin{eqnarray*}
&&%
\begin{tabular}{||l|l|l|l||}
\hline\hline
& $\alpha $ & $\eta $ & $\widehat{\eta }$ \\ \hline
BA graph & $1.96$ & $5.76$ & $5.14$ \\ \hline
generalized BA & $1.92$ & $5.60$ & $5.28$ \\ \hline
AS & $1.61$ & $4.21$ & $4.14$ \\ \hline
IG & $1.23$ & $2.74$ & $2.45$ \\ \hline\hline
\end{tabular}
\\
&&\text{Table 2. The tail exponents of networks}
\end{eqnarray*}%
The lobby index is placed on the map along other centrality measures. \ Some
further extensions and properties are discussed as well: the relation to the
diplomat's dilemma is investigated and the lobby index is demonstrated to be
a good performance measure for lobbyists.

\emph{Acknowledgments }We thank Tam\'{a}s Nepusz who kindly provided the
calculations of the classical centralities and their correlations in Table
1. for the AS graph and gave same valuable remarks on the manuscript. We
also thank F\"{u}l\"{o}p Bazs\'{o} for his useful comments on the draft.
Thanks are due to Tommaso Colozza and Erik Bodzs\'{a}r who gave very
valuable numerical calculations and simulation results for this work. \ The
authors were partially supported by the High Speed Network Laboratory of the
Budapest University of Technology and Economics. The authors also wish to
thank the referees for their helpful comments and suggestions.

\section{Appendix}

In what follows we provide a rigorous derivation of $\left( \ref{main}%
\right) $\footnote{%
Henceforth $c$ will be an arbitrary positive constant unless specified
otherwise. \ Its value may change from occurrence to occurrence.}. Let us
use the notation $l_{k}={\mathbb{P}}\left( l\left( x\right) =k\right) $ for
the distribution of the $l$-index, and $G_{k}={\mathbb{P}}\left( \deg \left(
x\right) \geq k\right) =1-F_{k}$. 
\begin{eqnarray*}
l_{k} &=&\sum_{l=0}^{\infty }{\mathbb{P}}\left( l\left( x\right) =k,\deg
\left( x\right) =k+l\right) \\
&=&\sum_{l=0}^{\infty }{\mathbb{P}}\left( l\left( x\right) =k|\deg \left(
x\right) =k+l\right) \mathbb{P}\left( \deg \left( x\right) =k+l\right)
\end{eqnarray*}%
Partition of unity and conditional probability is used (Bayes Theorem). As a
result we have to investigate what is the probability that a node has $k$
links, each has degree $\geq k$ and $l$ other links with degree not higher
then $k$ given that it has $k+l$ links in total.\ That criteria makes
exactly $l\left( x\right) =k.$

First we develop a lower estimate for $l_{k}$. 
\begin{equation*}
l_{k}\geq c_{1}^{k}k^{-\alpha k}\sum_{l=0}^{\infty }\left( k+l\right)
^{-\left( \alpha +1\right) }\binom{k+l}{l}\left( 1-c_{1}k^{-\alpha }\right)
^{l}
\end{equation*}%
We estimate $l_{k}$ using $1-c_{1}k^{-\alpha }\approx e^{-c_{1}k^{-\alpha }}$
and $\binom{k+l}{l}\geq \frac{l^{k}}{k!}$%
\begin{eqnarray*}
l_{k} &\geq &c\frac{c_{1}^{k}k^{-\alpha k}}{k!}\sum_{l=0}^{\infty
}l^{k-\left( \alpha +1\right) }e^{-\frac{c_{1}l}{k^{\alpha }}} \\
&\geq &c\frac{c_{1}^{k}k^{-\alpha k}}{k!}\int_{0}^{\infty }x^{k-\left(
\alpha +1\right) }e^{-\frac{c_{1}x}{k^{\alpha }}}dx \\
&\geq &cc_{1}^{k}k^{-\alpha k}\frac{\Gamma \left( k-\left( \alpha +1\right)
+1\right) }{k!}\left( \frac{k^{\alpha }}{c_{1}}\right) ^{k-\left( \alpha
+1\right) +1} \\
&=&ck^{-\alpha ^{2}}\frac{\Gamma \left( k-\alpha \right) }{k!} \\
&\geq &ck^{-\alpha ^{2}}\frac{\left( k-\alpha \right) ^{k-\alpha
-1/2}e^{k-\alpha +\Theta /12\left( k-\alpha \right) }}{k^{k+1/2}e^{k+\Theta
/12\left( k\right) }} \\
&\geq &ck^{-\alpha \left( \alpha +1\right) -1}.
\end{eqnarray*}%
where the Stirling formula has been used for $\Gamma \left( k-\alpha \right) 
$ and $k!$ as well and $0<\Theta <1$. The upper estimate works similarly as
follows.%
\begin{equation*}
l_{k}=\left( G_{k}\right) ^{k}\sum_{l=0}^{\infty }\left( k+l\right)
^{-\left( \alpha +1\right) }\binom{k+l}{l}\left( F_{k+1}\right) ^{l}.
\end{equation*}%
\begin{eqnarray*}
l_{k} &\leq &c\frac{c_{1}^{k}k^{-\alpha k}}{k!}\sum_{l=0}^{\infty }\left(
k+l\right) ^{k-\left( \alpha +1\right) }e^{-\frac{c_{1}l}{\left( k+1\right)
^{\alpha }}} \\
&\leq &c\frac{c_{1}^{k}k^{-\alpha k}}{k!}\sum_{l=0}^{\infty }\left(
k+l\right) ^{k-\left( \alpha +1\right) }e^{-\frac{c_{1}l}{k^{\alpha }}\left( 
\frac{k}{k+1}\right) ^{\alpha }} \\
&=&c\frac{c_{1}^{k}k^{-\alpha k}}{k!}\sum_{l=k}^{\infty }l^{k-\left( \alpha
+1\right) }e^{-\frac{c_{1}\left( l-k\right) }{k^{\alpha }}\left( \frac{k}{k+1%
}\right) ^{\alpha }} \\
&=&c\frac{c_{1}^{k}k^{-\alpha k}}{k!}\sum_{l=k}^{\infty }l^{k-\left( \alpha
+1\right) }e^{-\frac{c_{1}l}{k^{\alpha }}\left( \frac{k}{k+1}\right)
^{\alpha }} \\
&\leq &c\frac{c_{1}^{k}k^{-\alpha k}}{k!}\sum_{l=0}^{\infty }l^{k-\left(
\alpha +1\right) }e^{-\frac{c_{1}l}{k^{\alpha }}\left( \frac{k}{k+1}\right)
^{\alpha }} \\
&\leq &c\frac{c_{1}^{k}k^{-\alpha k}}{k!}\int_{0}^{\infty }x^{k-\left(
\alpha +1\right) }e^{-\frac{\left( \frac{k}{k+1}\right) ^{\alpha }c_{1}x}{%
k^{\alpha }}}dx
\end{eqnarray*}

Introducing a new variable one obtains%
\begin{eqnarray*}
&&l_{k}\leq c\frac{c_{1}^{k}k^{-\alpha k}}{k!}\int_{0}^{\infty }x^{k-\left(
\alpha +1\right) }e^{-\frac{\left( \frac{k}{k+1}\right) ^{\alpha }c_{1}x}{%
k^{\alpha }}}dx \\
&=&c\frac{c_{1}^{k}k^{-\alpha k}}{k!}\left[ \frac{\left( \frac{k}{k+1}%
\right) ^{\alpha }c_{1}}{k^{\alpha }}\right] ^{-k+\alpha }\int_{0}^{\infty
}y^{k-\left( \alpha +1\right) }e^{-y}dy \\
&=&cc_{1}^{k}k^{-\alpha k}\frac{\Gamma \left( k-\left( \alpha +1\right)
+1\right) }{k!}\left( \frac{k^{\alpha }}{c_{1}}\right) ^{k-\alpha }\left[
\left( \frac{k+1}{k}\right) ^{\alpha }\right] ^{k-\alpha } \\
&\leq &ck^{-\alpha ^{2}}\frac{\Gamma \left( k-\alpha \right) }{k!}\left( 1+%
\frac{1}{k}\right) ^{k} \\
&\leq &ck^{-\alpha ^{2}}\frac{\left( k-\alpha \right) ^{k-\alpha
-1/2}e^{k-\alpha +\Theta /12\left( k-\alpha \right) }}{k^{k+1/2}e^{k+\Theta
/12\left( k\right) }} \\
&\leq &ck^{-\alpha \left( \alpha +1\right) -1}
\end{eqnarray*}%
where at the end the Stirling formulas have been used as in the lower
estimate.

\bibliographystyle{iopart-num}
\bibliography{lindex}

\end{document}